\documentclass{aastex}
\usepackage{palatino}
\usepackage{bm}
\usepackage[papersize={7.2in,9.6in},vscale=0.88,hscale=0.81,includehead,vmarginratio=1:1]{geometry}
\usepackage{graphicx}
\usepackage{amssymb,amsbsy,amsmath}
\usepackage{hyperref}
\newcommand{\dem}{{\cal E}}
\newcommand{\demcoff}{{e}}
\newcommand{\dataelmt}{{I}}
\newlength{\columnheight}
\newlength{\figwidth}
\newlength{\ruleheight}
\begin{document}

\title{Fast Differential Emission Measure Inversion of Solar Coronal Data}
\author{Joseph Plowman, Charles Kankelborg, Petrus Martens}
\affil{Montana State University}
\affil{Bozeman, MT, USA}

\begin{abstract}
We present a fast method for reconstructing Differential Emission Measures (DEMs) using solar coronal data. On average, the method computes over 1000 DEMs per second for a sample active region observed by the Atmospheric Imaging Assembly (AIA) on the Solar Dynamics Observatory (SDO), and achieves reduced chi-squared of order unity with no negative emission in all but a few test cases. The high performance of this method is especially relevant in the context of AIA, which images of order one million solar pixels per second. This paper describes the method, analyzes its fidelity, compares its performance and results with other DEM methods, and applies it to an active region and loop observed by AIA and by the Extreme-ultraviolet Imaging Spectrometer (EIS) on Hinode. 
\end{abstract}

\noindent

\maketitle

\section{Introduction}

The Sun's outer atmosphere, the corona, is a hot ($\sim 10^6$ Kelvin), highly magnetized, dynamic plasma, with spatial scales ranging from the electron gyroradius of less than 1 cm to the solar radius of over $10^{13}$ cm. As such, it is one of the most fruitful objects for the study of astrophysical plasmas. One of the most fascinating puzzles of the Corona is how it is heated to temperatures $\sim 1000$ times those of the underlying photosphere. It is believed that this process is magnetic in nature, but the details (e.g., the spatial distribution of heating and whether it is continuous or impulsive) are not known. One important tool for understanding these details is the reconstruction of solar temperature distributions called `Differential Emission Measures' (DEMs) from spectral images of the Sun.

The DEM, $\dem (T)$ characterizes coronal emission at a given temperature\footnote{For most spectral lines; lines corresponding to forbidden transitions have more complicated density sensitivity, but that is beyond the scope of this paper. Interested readers may consult \citet{delzannamasonasr2005}.}, and is defined in terms of the densities, $n(l)$, along the line-of-sight so that $\int\dem (T) dT \equiv \int n^2(l)dl$ (column emission convention). Observed intensities, $\dataelmt_n$,
are given by integrating the instrument response, $R_n(T)$, against the DEM being observed:
\begin{equation}\label{eq:demconvolution}
	\dataelmt_n = \int R_n(T)\dem (T)dT
\end{equation}

Algorithms which reconstruct these temperature distributions must be very fast if they are to process the volume of data produced by modern solar observatories, or resolve the dynamics and fine spatial scales believed to be involved in coronal heating. For example, matching the real-time data rate of AIA \citep{lemenetal_aia_soph_2012}, requires computation of over $10^5$ DEMs per second.  
There are a number of existing algorithms for reconstructing DEMs from solar image data, but they are far too slow to meet this requirement. 
One widely used method is the PINTofALE code \citep{kashyap_drake_apj_1998}, which employs a Markov Chain Monte Carlo (MCMC) search taking seconds or minutes to compute a DEM. Another method, which is considered relatively fast and was recently applied to AIA data by \citet{hannahkontaraap2012}, computes $\sim 4$ DEMs per second (an updated version with substantially improved performance has recently been made available at \url{http://www.astro.gla.ac.uk/~iain/demreg/map/}; we use this version in our comparisons).

We present a fast, iterative, regularized method for inferring DEMs using data from solar imagers such AIA and EIS. With one thread on a 3.2GHz processor, it is able to compute well over 1000 DEMs per second for an example solar active region observed by AIA, and over $100$ DEMs per second in our most difficult test cases. Moreover, we anticipate that with straightforward optimizations (e.g., conversion of computationally intensive portions of the code to C and parallelization), the performance of the code will be increased to $\sim 10^5$ DEMs per second on a single workstation, sufficient to match AIA's real-time observing rate. 

This paper describes the method, analyzes its fidelity, compares its performance and results with other DEM methods, and applies it to example solar data. We also touch on the limitations of the solar data to constrain the DEM being observed. It must be noted that the ability to recover the details of the original solar input DEM is limited, both because of the number of channels are limited and because the temperature response functions are broad, due to the width of the temperature-dependent emissivities of the spectral lines from which they are constructed. This is discussed in detail, for instance, in \citet{craigbrownaap1976,judgeetalapj1997}.

\section{DEM Algorithm}
\subsection{First Pass}\label{sub:firstpass}
We wish to infer the DEM, a continuous function, from a small number of observed intensities which are the result of the convolution of the DEM with the instrument response functions. This problem is inherently ill-posed due to the limited number of response functions and the loss of information resulting from the convolution (i.e., equation \ref{eq:demconvolution}) . To resolve the ambiguity, we impose additional constraints on the DEM solution. The first is that the DEM can be expressed as a linear combination of some set of basis elements (e.g., a set of narrow temperature bins), $B_j$, with coefficients $e_j$ (i.e., $\dem (T) = \sum_j e_j B_j$). The integral equation (\ref{eq:demconvolution}) then becomes a matrix equation:
\begin{equation}\label{eq:demmatrix}
	\dataelmt_n = \int R_n(T) \dem (T)dT = \sum_k\Big[ \demcoff_k \int R_n(T) B_k(T)dT\Big] \equiv \sum_k \demcoff_k \gamma_n A_{nk},
\end{equation}
Where $\gamma_n$ are a set of normalization constants for the response functions, which we choose to be their squared integral, square root. 

If the number of basis elements is greater than number of instrument channels, the underconstrained inversion can be resolved by a Singular Value Decomposition \citep[SVD;][]{nr}, which picks the coefficient vector with the smallest magnitude (i.e., least squared emission measure). This is a sort of smoothness constraint, since a solution which has emission concentrated into narrow peaks will have greater total squared EM than a solution which spreads the emission as broadly as the data will allow. Such a solution will also tend to reduce non-physical negative emission, since negative EM is likely to require excess positive emission measure elsewhere in order to produce the observed intensities, resulting in relatively high total squared EM.

If the number of basis elements is equal to the number of instrument channels, it is straightforward to invert equation \ref{eq:demmatrix} to find the DEM coefficients, $\demcoff_k$. In that case, however, the quality of the inversion is highly sensitive to choice of basis. In particular, many choices of basis lead to an inversion matrix (i.e., $A_{jk}^{-1}$) that has large negative entries, causing large negative values in the resulting DEMs. 

The tendency of the inverse matrix to have large negative values is reduced, however, if $A_{jk}$ is diagonally dominant (i.e., each diagonal entry of the matrix is the largest one in it's respective row/column). In particular, if we choose the basis functions to be the instrument response functions themselves, square-normalized, we obtain a symmetric $A_{jk}$ matrix whose diagonal entries are unity, with all off-diagonal entries less than one.
Remarkably, we find that this basis gives results identical to using a large number of narrow basis elements and an SVD\footnote{Appendix \ref{app:instresp_demderivation} demonstrates why this is the case}. It therefore satisfies the SVD's minimum squared emission measure constraint, and we use it to compute a first pass DEM as a starting points for our DEM inversion. It reduces the number of operations required for each pixel to only $\sim N^2$ operations with $N$ instrument channels, increasing the speed of the inversion. 

Even when using the response functions as a basis, however, we continue to compute $A_{ij}^{-1}$ using an SVD. This allows us to enforce a minimum condition ($\sim 10^{-12}$, typically\footnote{This means the smallest singular value used to form the inverse matrix must be at least $10^{-12}$ times the largest singular value.}) on the inversion, ensuring that round-off error (from $A_{ij}$ being nearly singular) does not cause ringing. In effect, this combines highly similar channels (or linear combinations of channels, to be more precise) into a single channel rather than trying to fit them individually. 

\subsection{Regularization}\label{sub:reg}
Another constraint which DEM solutions must satisfy is a physically required positivity constraint. However, the first pass DEM inversion described in Section \ref{sub:firstpass} can produce negative coefficients and therefore negative EM. This is particularly true when the errors in the observed intensities are significant, as the first-pass solutions exactly reproduce the input data and there is no guarantee that a set of noisy data intensities can be exactly reproduced by a purely positive DEM. To mitigate this issue, we once again seek to minimize the total squared emission measure, but we now allow the DEM to deviate from the input data at a specified $\chi^2$ level. This is implemented by seeking a new DEM which minimizes the sum
\begin{equation}
	\chi^2+\lambda\int [\dem (T)]^2dT = \sum_j\frac{\Delta \dataelmt_j}{\sigma_j^2} +  \lambda\sum_{jk}\demcoff_j \demcoff_k A_{jk},
\end{equation}
where $\dem (T)$ is the DEM described above, and $\lambda$ is a regularization parameter chosen to enforce the desired $\chi^2$ threshold, $\chi^2_0$. Using our basic DEM solution, we replace $\demcoff_i$ with a set of corresponding regularization corrections to the data values, $\Delta \dataelmt_i$, so that $\demcoff_j = \sum_k A^{-1}_{jk} (\dataelmt_k+\Delta \dataelmt_k)/\gamma_k$:
\begin{equation}
	\chi^2+\lambda\int [\dem (T)]^2dT = \sum_j\frac{\Delta \dataelmt_j^2}{\sigma_j^2} + \lambda\sum_{jk}(\dataelmt_j+\Delta \dataelmt_j)\frac{A_{jk}^{-1}}{\gamma_j\gamma_k}(\dataelmt_k+\Delta \dataelmt_k).
\end{equation}
This nonlinear system of equations may be solved for the data corrections, $\Delta\dataelmt_j^2$, and regularization parameter $\lambda$, satisfying the desired $\chi^2$ threshold $\chi_0^2$, in a variety of ways; we have found that a standard bisection search \citep[see][for example]{nr} gives acceptable performance. The new regularized DEM is simply the first-pass inversion applied to the new corrected data values.

Regularized solution of the DEM problem has recently been discussed by \citet{hannahkontaraap2012}, but our regularization is simpler and faster owing to the basis set used for the DEM. Since DEMs constructed from the instrument response functions have only a small number of basis elements, however, they remain liable to producing negative emission in cases with sharp features. The solution to this problem is discussed next.

\subsection{Enforcing Non-Negativity}
We remove the remaining negative emission via the following iterative process: 
\begin{enumerate}
	\item Zero the negative EM in the current DEM, $\dem^{(n)}$, to create a new DEM, $\dem_+^{(n)}$. At the zeroth iteration, this is the regularized DEM from Section \ref{sub:reg} $\dem^{(0)}=\sum_j \demcoff_j B_j(T)=\sum_{jk} B_j(T) A^{-1}_{jk} (\dataelmt_k+\Delta \dataelmt_k)/\gamma_k$.\label{enum:initialstep}
	\item Compute the data intensities, $\dataelmt_j^+=\int \dem_+^{(n)}(T)R_j(T)dT$, corresponding to $\dem_+^{(n)}$.\label{enum:iterationposintensities}
	\item Take the difference between $\dataelmt_j^+$ and the original $\dataelmt_j$, $\Delta \dataelmt_j^+ = \dataelmt_j^+-\dataelmt_j$.
	\item Compute correction DEM coefficients, $\Delta \demcoff_j = \sum_k A_{jk}^{-1}\Delta \dataelmt_k^+/\gamma_k$. $A_{jk}^{-1}$ is computed using an SVD at this step, and we enforce a relatively strong minimum condition number ($\sim 0.1$) to reduce ringing in the correction.\label{enum:iterationcorrectionstep}
	\item Subtract the corresponding DEM corrections, $\Delta \dem^{(n)}$ from $\dem_+^{(n)}$. By construction, this restores $\dem^{(n+1)} \equiv \dem_+^{(n)}-\Delta \dem^{(n)}$ to agreement with 
	the data, but reintroduces some negative emission.
	\item Repeat from step \ref{enum:initialstep} until $\dataelmt_i^+$ matches $\dataelmt_i$ to within the desired $\chi^2$.
\end{enumerate}

After we zero the negative EM in the initial DEM (i.e., at step \ref{enum:initialstep} of the zeroth iteration), the DEM is no longer represented by a basis of instrument response functions, but rather by a continuous function (see step \ref{enum:iterationposintensities} of the iteration). In practice, we choose to express it using an intermediate basis of closely spaced, narrow functions - usually triangle functions. 

To further speed convergence of the iteration, we attempt a linear extrapolation at each step of the iteration. The extrapolation steps move the DEM vector further along the direction of the current iteration step, and are only accepted when they improve $\chi^2$. 

We find that the optimal regularization strength, and minimum condition number strength in step \ref{enum:iterationcorrectionstep} of the iteration, vary with input DEM and data quality, so we try multiple values of these parameters, beginning with a light regularization and small minimum condition number threshold and moving to stronger regularization and larger minimum condition number threshold. Typical values of these parameters are $[0.9,0.5,0.01]$\footnote{We define the regularization parameters to be p-values of the $\chi^2$ distribution \citep[see, for instance, Chapter 15 of][where they are referred to as $Q$, and $P$ refers to their complement]{nr}. With a regularization parameter of 0.9, for instance, the $\chi^2$ of the regularized data with respect to the original data will correspond to a p-value of $0.9$.} and $[0.01,0.05,0.1]$, for the regularization strengths and minimum condition numbers, respectively. In broad terms, the operation of the algorithm can then be described as follows:
\begin{itemize}
\item Apply the light (e.g., 0.9) regularization to the data and compute the corresponding first-pass DEM.
\item Zero the negative emission in the first-pass DEM and compute the $\chi^2$ of the resulting data with respect to the initial data. If $\chi^2$ exceeds the desired threshold, attempt to iterate away the negative emission, using the smallest minimum condition threshold (e.g., 0.01), until the $\chi^2$ threshold is reached.
\item If the iteration takes too long to reach the desired $\chi^2$ threshold, retry the first two steps with the next strongest regularization strengths and minimum conditions (e.g., 0.5 and 0.05, respectively). Repeat until the $\chi^2$ threshold is reached, or each pair of regularization strengths and minimum conditions have been tried.
\end{itemize}

Readers who would like more implementation details are encouraged to examine our code, which we have made publicly available at \url{http://solar.physics.montana.edu/plowman/firdems.tgz}. We also intend to submit the code to SolarSoft in the near future.

\section{Fidelity \& Performance}
We have tested our DEM using a variety of example cases. The results are given in figures on the following pages, which show recovered DEMs, $\chi^2$ agreement with the data, and computational time. Please note the following:
\begin{itemize}
\item All reported $\chi^2$ have been reduced by dividing by the number of instrument channels being fit. 
\item Unless otherwise noted, we use a one second exposure time for AIA, and 30 seconds for EIS. EIS pixels were assumed to be sampled to 2 square arcseconds.
\item AIA errors are assumed to be from read and shot noise alone, while EIS errors assume an
additional 10\% error from other sources.
\item We use the 94, 131, 171, 193, 211, and 335 \AA\ AIA channels throughout. 
\item EIS emissivities were computed assuming a density of $10^9 \ \mathrm{cm}^{-3}$, unless otherwise noted.
\item AIA response functions were computed by calling \texttt{aia\_get\_response(/temp, /dn, /evenorm)}. 
\item Computation times are for a single thread running on a 3.2GHz Intel Xeon processor, in IDL. 
\end{itemize}

Wherever possible, we use $\log_{10}(T)$ as a temperature variable rather than $T$ (in Kelvin) itself. We believe this is the more natural parameter for the temperature, since we are interested in a temperature range spanning multiple orders of magnitude ($\sim 10^5 - 10^7$ Kelvin). Similarly, our DEMs are scaled per unit $\log_{10}(T)$, rather than per unit $T$. We believe this is the natural scaling when the DEMs are represented as functions of $\log_{10}(T)$, because the area under the DEM curves is then the emission measure. Rescaling to unit $T$ may be accomplished by dividing by $T\ln{(10)}$.

\begin{figure}[!ht]
	\caption{DEM inversions of Log-normal simulated DEMs of width 0.2 using the six AIA EUV channels. The input DEM is shown by the solid line, while each dotted line is a DEM inversion with randomly chosen read and shot noise.}\label{fig:aia_lognormaltest01_5e28}
	\begin{center}\includegraphics[width=\figwidth]{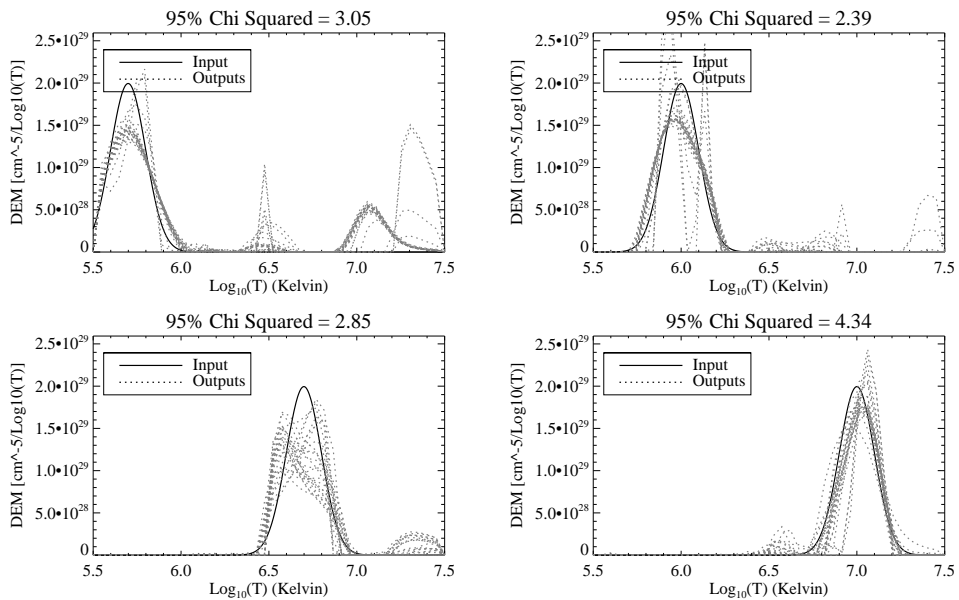}\end{center}
\end{figure}

\begin{figure}[!ht]
	\caption{Same as figure \ref{fig:aia_lognormaltest01_5e28}, but for the 24 EIS lines used in \citet{warrenetal_apj_2011}}\label{fig:eis_lognormaltest01_5e28}
	\begin{center}\includegraphics[width=\figwidth]{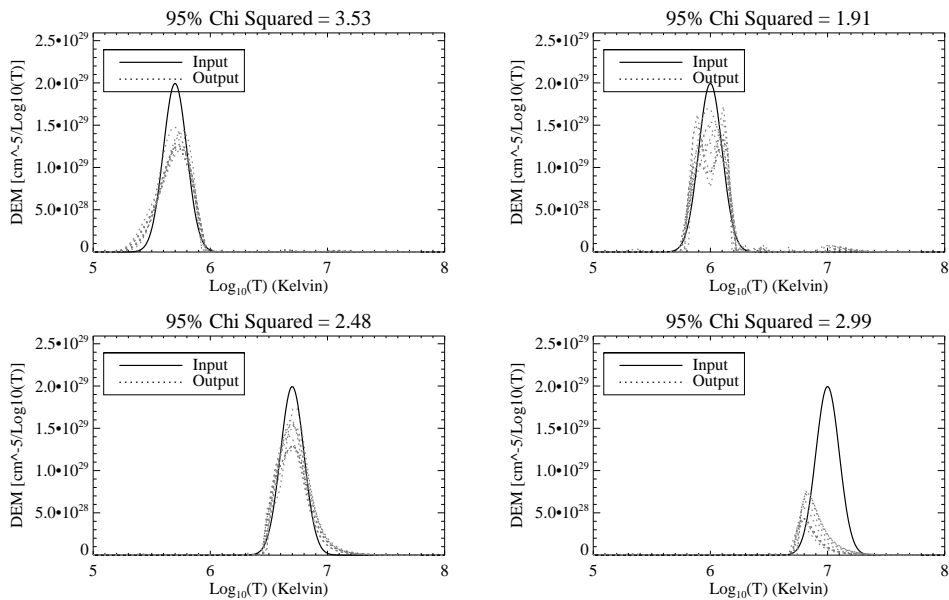}\end{center}
\end{figure}

\begin{figure}[!ht]
	\caption{AIA Response  to Log-normal DEMs of width 0.2 and total EM $5.0 \times 10^{28} \textrm{cm}^{-5}$ at temperatures from 5.5 to 7.0 dex. Left: the recovered DEMs - each vertical slice of this plot is a DEM like one of the output curves of Fig. \ref{fig:aia_lognormaltest01_5e28}, at the temperature of the corresponding x axis value. The solid lines on left show emission measure weighted median temperature (EMWMT). Center: $\chi^2$ percentiles resulting from repeated MC trials of the read and shot noise, at each temperature. Right: The average time per DEM at each temperature.}\label{fig:aia_gresp_02_5e28}
	\begin{center}\includegraphics[width=\figwidth]{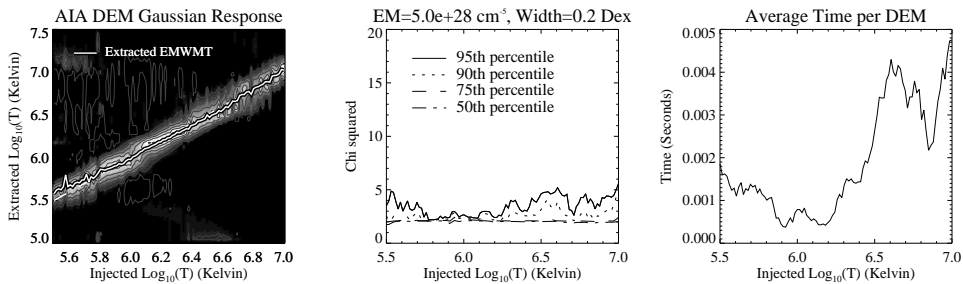}\end{center}
\end{figure}

\begin{figure}[!ht]
	\caption{Same as Fig. \ref{fig:aia_gresp_02_5e28}, but for Log-normal DEMs of width 0.3.}\label{fig:aia_gresp_03_5e28}
	\begin{center}\includegraphics[width=\figwidth]{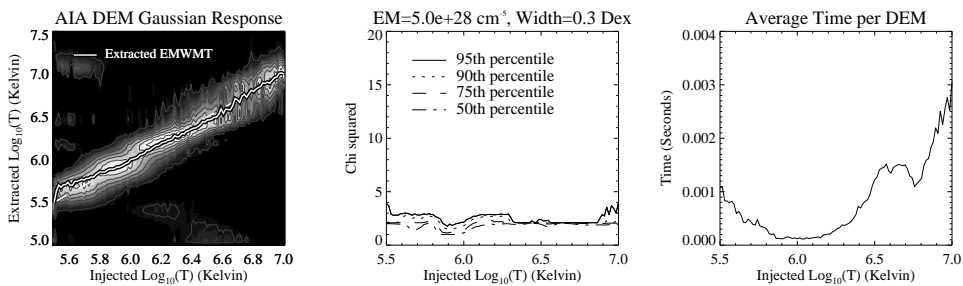}\end{center}
\end{figure}

\begin{figure}[!ht]
	\caption{EIS Response  to Log-normal DEMs of width 0.2 and total EM $5.0 \times 10^{28}\textrm{cm}^{-5}$ at temperatures from 5.5 to 7.0 dex. Solid lines on left shows emission measure weighted median temperature (EMWMT). The 24 spectral lines from \citet{warrenetal_apj_2011} were used.}\label{fig:eis_gresp_02_5e28}
	\begin{center}\includegraphics[width=\figwidth]{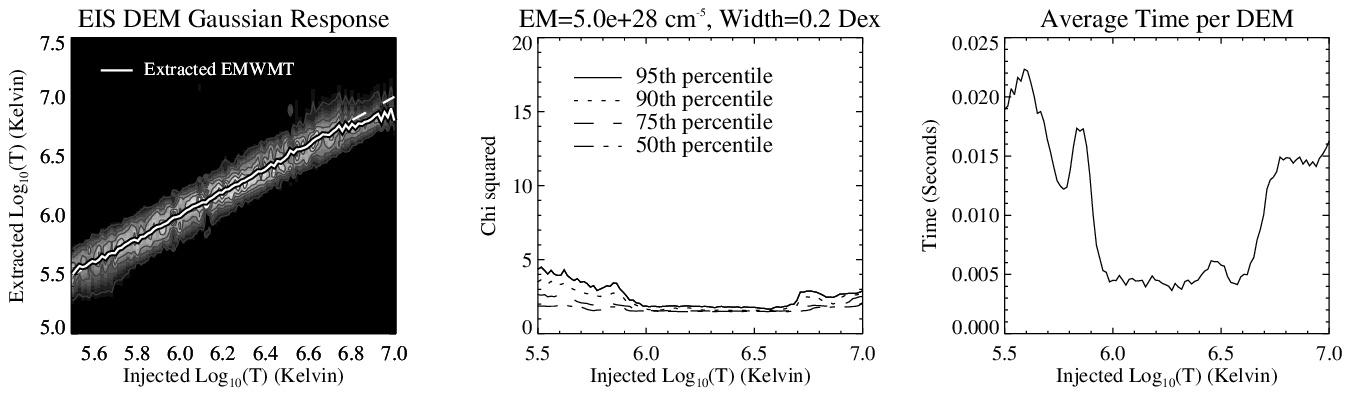}\end{center}
\end{figure}

\begin{figure}[!ht]
	\caption{EIS Response  to Log-normal DEMs of width 0.3 and total EM $5.0 \times 10^{28} \textrm{cm}^{-5}$ at temperatures from 5.5 to 7.0 dex. Solid lines on left shows emission measure weighted median temperature (EMWMT). The 24 spectral lines from \citet{warrenetal_apj_2011} were used.}\label{fig:eis_gresp_03_5e28}
	\begin{center}\includegraphics[width=\figwidth]{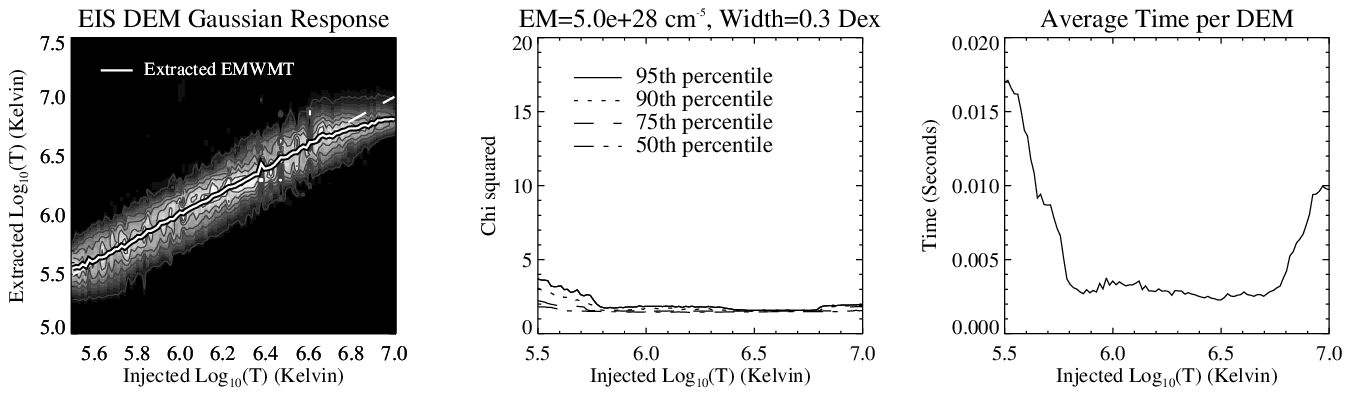}\end{center}
\end{figure}

\begin{figure}[!ht]
	\caption{AIA inversion of DEMs produced by summing five Log-normal DEMs with randomly chosen centers, widths, and amplitudes. Center locations are uniformly distributed between $10^{5.75}$ and $10^7$, widths between $0.1$ and $0.3$ and total EM between $5\times 10^{27} \textrm{cm}^{-5}$ and $5\times 10^{28} \textrm{cm}^{-5}$. Top: Results from our fast DEM Method. Bottom: Same for \citet{hannahkontaraap2012} regularized DEM method, without positivity constraint.}\label{fig:aia_multimod}
	\begin{center}\includegraphics[width=\figwidth]{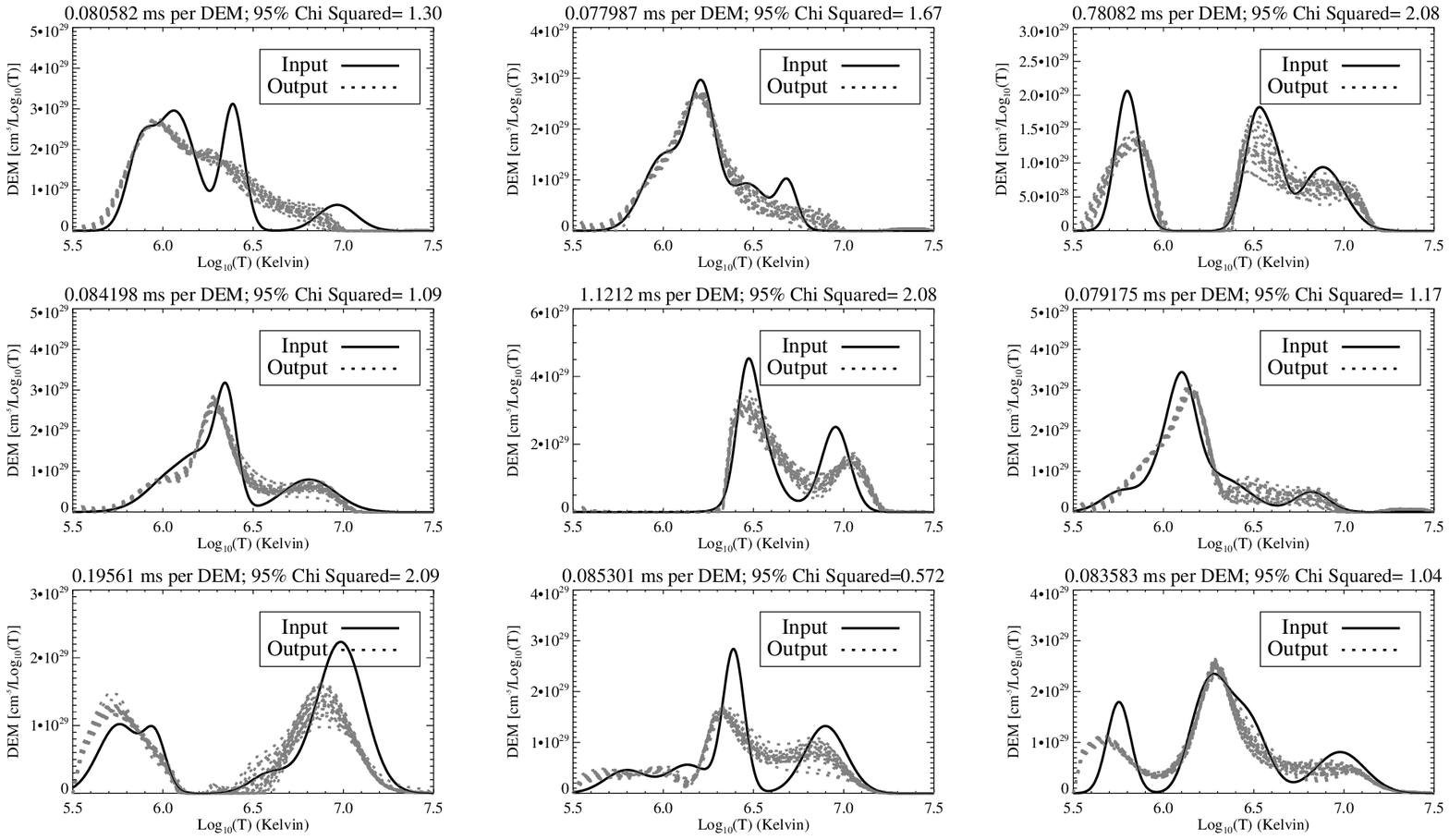}\end{center}

	\hrule
	\begin{center}\includegraphics[width=\figwidth]{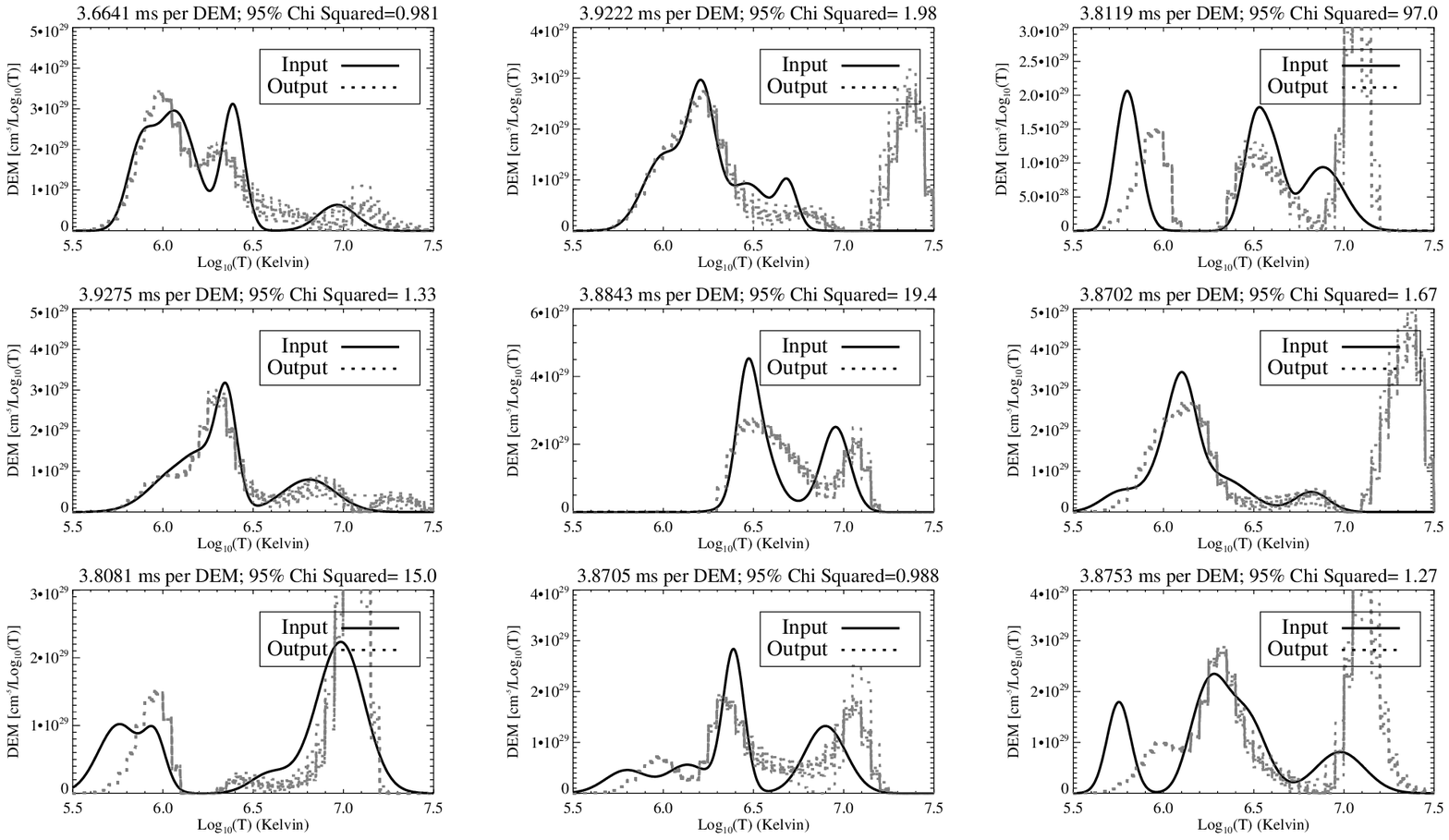}\end{center}
\end{figure}

\begin{figure}[!ht]
	\caption{Comparison of our DEM results (top) with \citet{warrenetal_apj_2011} (Figure 4) MCMC DEM results (bottom). We include their factor of 1.7 tweak to the coronal Mg abundance. The DEMs matched the data with $\chi^2$ of order unity.}\label{fig:warrendem_comparison}	\begin{center}\includegraphics[width=\figwidth]{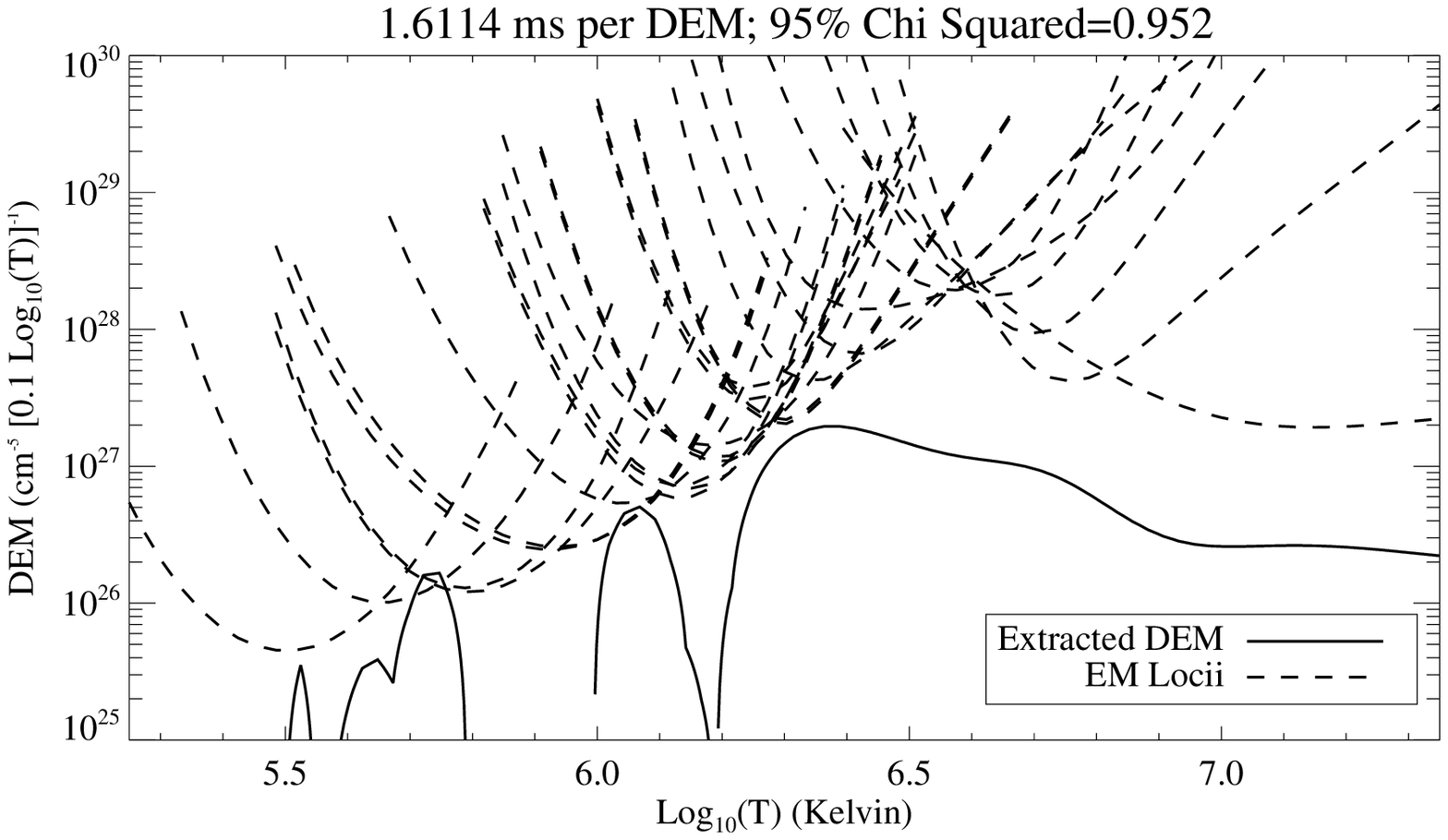}\end{center}
	\begin{center}\includegraphics[width=\figwidth]{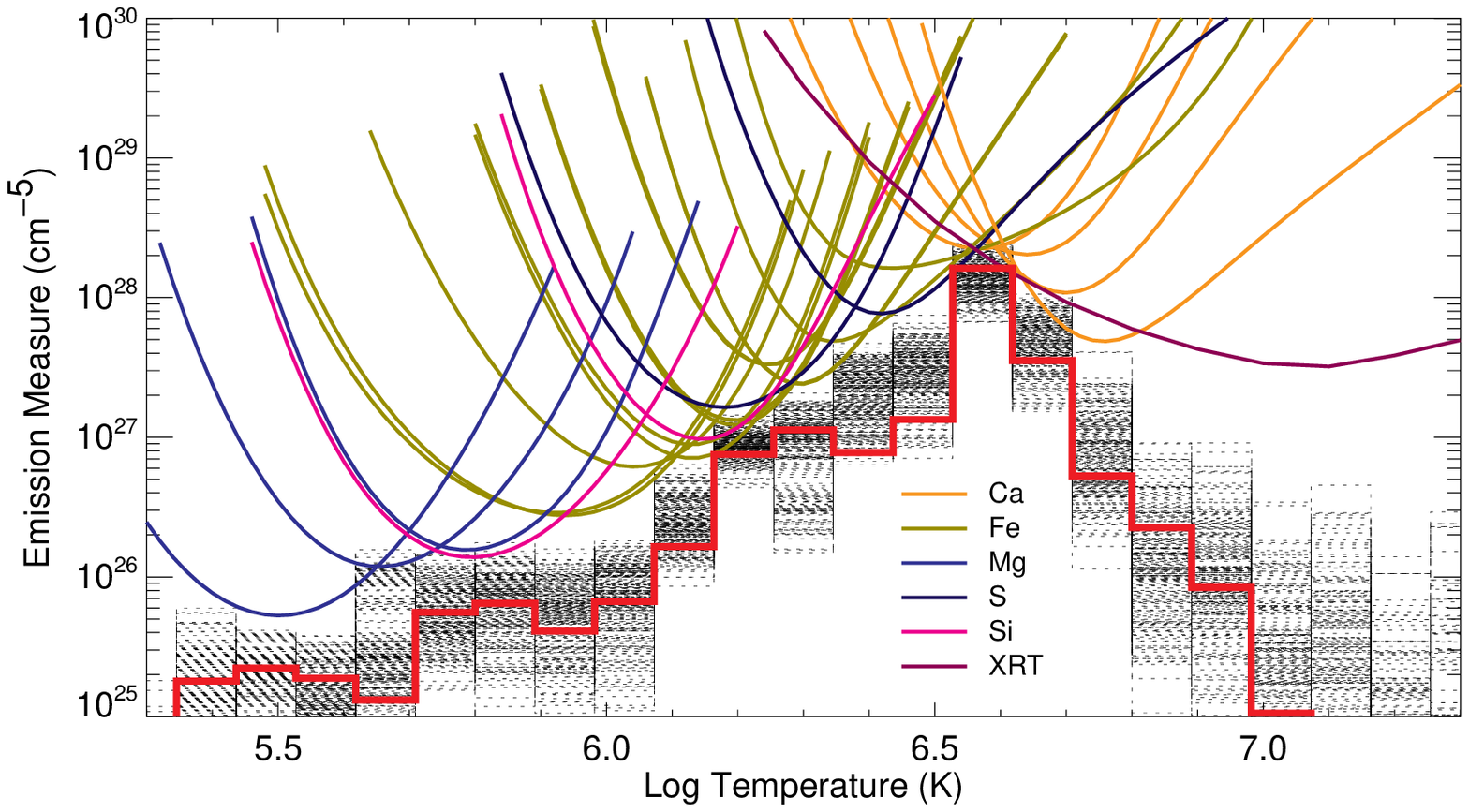}\end{center}
\end{figure}

With few exceptions, we find that we are able to recover the test cases with good $\chi^2$ (Reduced $\chi^2$ of one or two), that the recovered DEMs are a reasonable qualitative match to the input DEMs. Typical times for AIA DEMs are approximately one millisecond, with some cases taking under 0.1 millisecond. The test cases are as follows:
\begin{itemize}
	\item AIA inversion of Log-normal DEMs with widths of 0.2 at selected temperatures (Fig. \ref{fig:aia_lognormaltest01_5e28}).
	\item EIS inversion of Log-normal DEMs with widths of 0.2 at selected temperatures (Fig. \ref{fig:eis_lognormaltest01_5e28}).
	\item AIA inversion of Log-normal DEMs at temperatures ranging from $10^{5.5}$ to $10^{7.0}$ Kelvin, for widths of 0.2 (Fig. \ref{fig:aia_gresp_02_5e28}) and 0.3 (Fig. \ref{fig:aia_gresp_03_5e28}).
	\item EIS inversion (using the 24 lines chosen by Warren et al.\cite{warrenetal_apj_2011}) of Log-normal DEMs at temperatures ranging from $10^{5.5}$ to $10^{7.0}$ Kelvin, for widths of 0.2 (Fig. \ref{fig:eis_gresp_02_5e28}) and 0.3 (Fig. \ref{fig:eis_gresp_03_5e28}).
	\item AIA inversion of DEMs produced by summing five Log-normal DEMs with randomly chosen centers, widths, and amplitudes (Fig. \ref{fig:aia_multimod}).
	\item EIS Active region DEM from \cite{warrenetal_apj_2011} (Fig. \ref{fig:warrendem_comparison}). A density of $10^{9.5} \ \mathrm{cm}^{-5}$ is assumed, to match their emissivities.
\end{itemize}

%

We find that narrow (compared with the temperature response functions in question) DEMs are the most difficult for our method to reconstruct, in terms of obtaining good $\chi^2$ (i.e., $\chi^2_R \sim 1$) without negative emission. This can be seen by comparing Figures \ref{fig:aia_gresp_02_5e28} and \ref{fig:aia_gresp_03_5e28} (for AIA), or Figures \ref{fig:eis_gresp_02_5e28} and \ref{fig:eis_gresp_03_5e28} (for EIS). For AIA, there is somewhat more difficulty in recovering narrow DEMs at temperatures above $10^{6.5}$ Kelvin, but even in that case, acceptable $\chi^2$ were achieved for the majority of noise realizations. Despite their relatively poor $\chi^2$, these reconstructed DEMs are well-behaved and localized at the center of the injected DEM; the median $\log_{10}(T)$ is recovered to within $0.1$ between $\log_{10}(T) \approx 5.8$ and $7.0$. The difficulty in achieving good $\chi^2$ is absent in cases where there the emission is not concentrated near a single temperature, as can be seen in figure \ref{fig:aia_multimod}. In all cases considered, the average time to compute a DEM was under 10 milliseconds, and less than 0.1 millisecond for some cases.

In the case of EIS, temperatures greater than $10^{6.8}$ Kelvin, above the highest temperature peak of the spectral lines used, are the most challenging. The median temperatures are accurately (i.e.,  $\delta\log_{10}(T)\lesssim 0.1$) recovered over a range of $\log_{10}(T)$, $5.6\dots 6.8$. Due the larger number of channels, the computational time is somewhat longer, at $\sim 10$ milliseconds for a narrow log-normal distribution. 

Figure \ref{fig:aia_multimod} compares our DEM inversions (top set) with those of the optimized \citet{hannahkontaraap2012} (bottom set) `demmap' code. We ran the their code without its positivity constraint option, finding better $\chi^2$ were achieved by simply zeroing the negative emission. The tests DEMs are randomly generated multimodal distributions, as observed by AIA. Both codes were run in a single thread on the 3.2 GHz processor mentioned above. The results are qualitatively quite similar, although the Hannah \& Kontar method appears to produce some spurious high-temperature emission. For these test cases, our DEM method is faster at $\sim 10^{-4}$ seconds for most DEMs compared with $3.7\times 10^{-3}$ seconds (without positivity constraint) for the Hannah \& Kontar DEM. Our method also appears to produce reasonable $\chi^2$ with zero negative emission more consistently than the Hannah \& Kontar DEM inversions.

%

Figure \ref{fig:warrendem_comparison} compares our DEM results with the PINTofALE MCMC results reported in Figure 4 of \citet{warrenetal_apj_2011}. These results include all 24 EIS lines used by Warren et al, with their factor of 1.7 tweak to the Mg intensities, along with the XRT Open/Al-thick filter. Our DEM curve is considerably different than that of \citet{warrenetal_apj_2011}. In particular, we do not find a peak in the DEM at $\log_{10}(T)\approx 6.6$. Despite this difference, we obtain a very reasonable $\chi^2$ of 1.16, which is fully consistent with the data. This suggests that the \citet{warrenetal_apj_2011} reconstruction may not represent the full range of DEMs consistent with their data.

\subsection{Example Solar DEM Analysis}

\begin{figure}[!ht]
	\begin{center}\includegraphics[width=\figwidth]{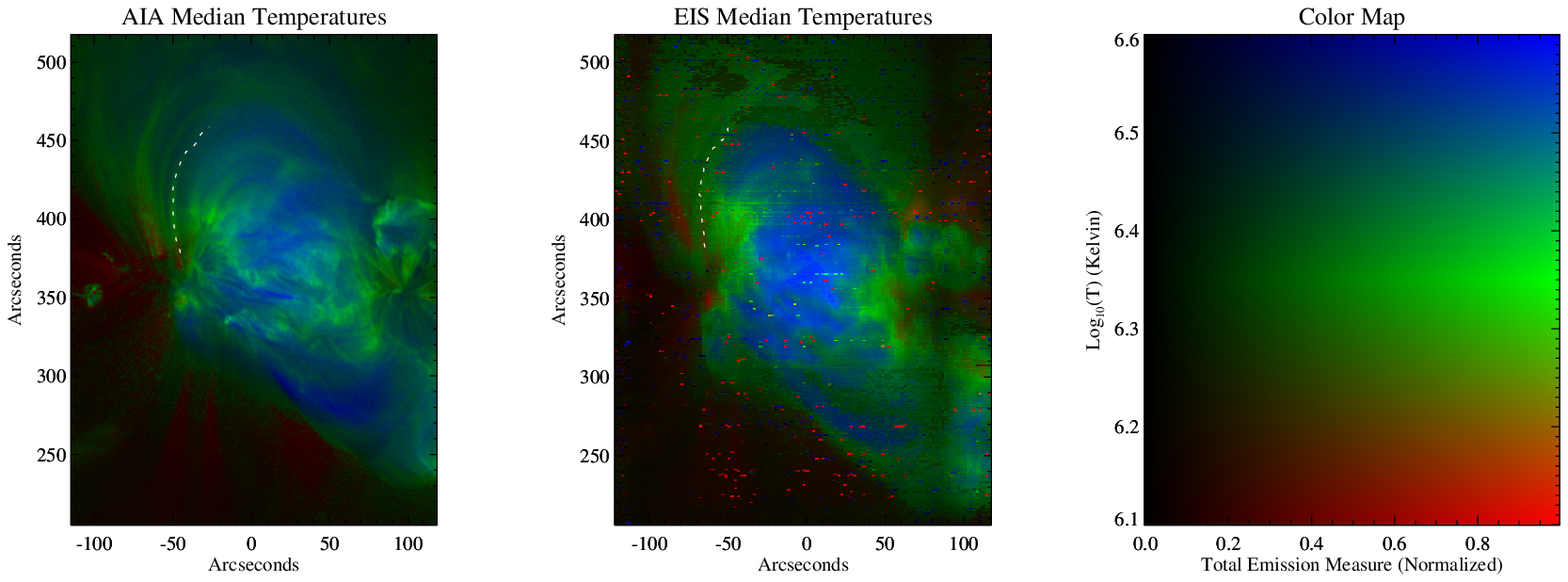}\end{center}
	\caption{Left: EMWMT (hue) and Total EM (intensity) from AIA DEM inversion of Active region (covered by EIS fov on April 19, 2011). Middle: Same for EIS. Right: color scale for  middle and left plots.}\label{fig:example_ardem}
\end{figure}

Finally, we show example DEM analysis of solar data. Figure \ref{fig:example_ardem} shows emission weighted median temperature (EMWMT) and total emission measure for an active region observed by EIS and AIA. The data cover the area of the large raster taken by EIS at around 01:30 on April 19, 2011. They are centered $\sim 350$ arcseconds above disk center, with an approximately $230\times510$ arcsecond field of view. The EIS data used fits to a set of five iron lines: Fe IX 188.497, Fe X 184.537, Fe XII 195.119, Fe XV 284.163, and Fe XVI 262.976. 

The average time per AIA DEM was approximately $0.13$ milliseconds, and we achieve reasonable $\chi^2$ for over 95\% of the AIA DEMs (the 95th percentile $\chi^2$ for EIS is 4.7).  The active region plots in Figure \ref{fig:example_ardem} are qualitatively quite similar, suggesting that the results give useful insight into the underlying solar temperature distribution. 

\begin{figure}[!ht]
	\begin{center}\includegraphics[width=\figwidth]{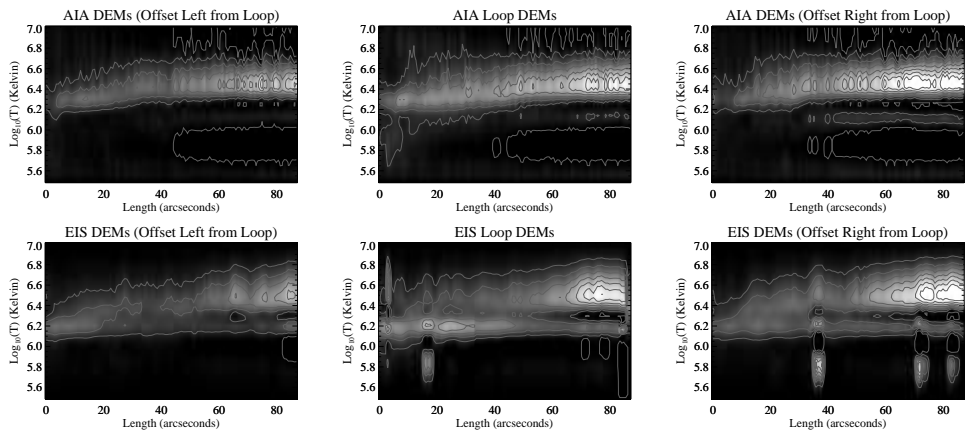}\end{center}
	\caption{Top: Loop DEM from AIA inversion of Active region (Center), offset 5 arcseconds to the left (left), 5 arcseconds to the right (right). Loop area indicated by dashed line in Figure \ref{fig:example_ardem}, or in more detail in Figure \ref{fig:aia_extracted_loop}. Bottom: Same for EIS.}\label{fig:example_loopdem}
\end{figure}

\begin{figure}[!ht]
	\begin{center}\includegraphics[width=\figwidth]{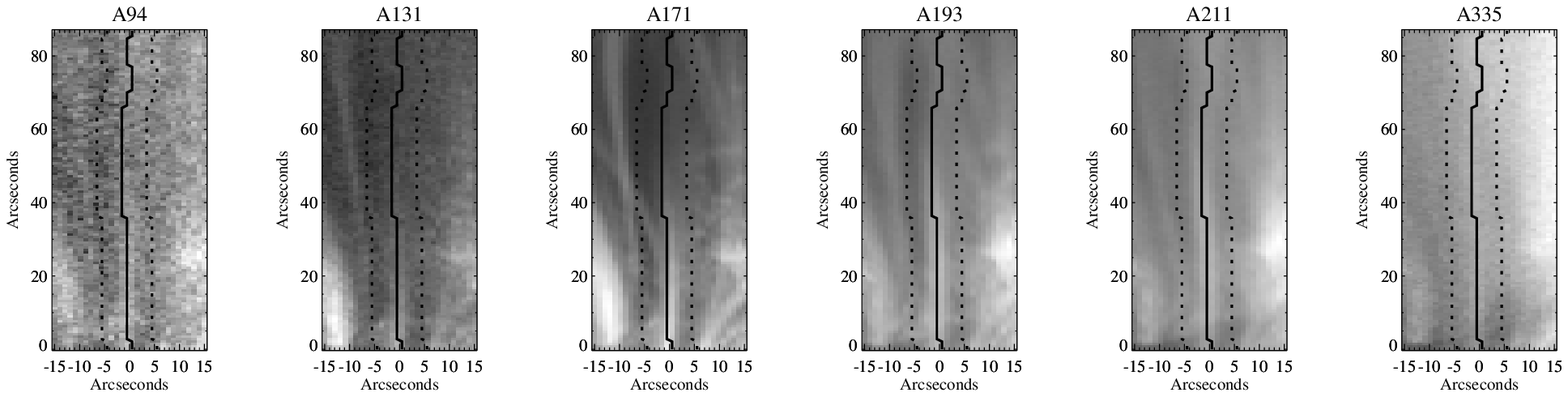}\end{center}
	\caption{Detailed, straightened AIA images of loop outlined in Figure \ref{fig:example_ardem}, for comparison with Figure \ref{fig:example_loopdem}. Loop trace indicated by solid line, while $\pm 5$ arcsecond offset background traces are indicated by dotted lines.}\label{fig:aia_extracted_loop}
\end{figure}

We also show DEMs along the loop segment indicated by the dashed line in Figure \ref{fig:example_ardem}. Figure \ref{fig:example_loopdem} shows DEMs along the length of the loop segment (solid lines in Figure \ref{fig:aia_extracted_loop}), as well as equivalent sets of DEMs offset to either side of the loop (dotted lines in Figure \ref{fig:aia_extracted_loop}). Once again, the DEMs found by AIA and by EIS are reasonably consistent, and both EIS and AIA inversions achieved good $\chi^2$ for over 95\% of the points along the loop. 

The DEMs shown along the loop axis are not made from background subtracted data, and they show little difference from the offset DEMs, which indicates that the loop emission has been swamped out be the bright background. In an upcoming work, we compute background subtracted DEMs for this loop and compare its temperature and density profile to a set of analytic strand heating models.

\section{Conclusions}

We have demonstrated a method for fast reconstruction of DEM distributions using coronal data from instruments such as EIS and AIA. This DEM method achieves reduced chi-squared of order unity with no negative emission in all but a few test cases. The most difficult test cases are narrow DEMs at high ($>10^{6.5}$ Kelvin) temperatures for AIA, and temperature regions with little spectral coverage for EIS. Even for the high temperature AIA cases, we achieve reasonable $\chi^2$ for the majority ($\sim 75\%$) of noise realizations. Qualitatively, the reconstructed DEMs match the input DEMs well, although the ability to recover finer details of the input DEMs is inherently limited. The data, particularly for AIA, do not constrain the fine details of the DEMs. When interpreting DEMs, great care must be exercised to determine whether or not the features of interest are genuine aspects of the coronal temperature distribution.

Our DEM method is fast enough to study the dynamics of the coronal temperature distributions in real time. For AIA data, the worst cases considered take under $10$ milliseconds, and some test cases execute in less than $0.1$ milliseconds. For the whole disk, at full AIA resolution using all six coronal EUV channels, we expect less than $\sim 1$ hour with the code as currently implemented. We plan to convert the computationally intensive parts of the code to C and rewrite them to take advantage of multithreading, which should offer a factor of $\sim 100$ performance gain on an eight core workstation. This would reduce the time for a set of full-resolution, full-disk AIA DEMs to under $\sim 1$ minute, and achieve the objective of matching the AIA observing rate in real time. The software is available online at \url{http://solar.physics.montana.edu/plowman/firdems.tgz}, and will be submitted to SolarSoft in the near future.

We also applied our DEM reconstruction to solar data, an active region observed by EIS and AIA on April 19, 2011. We found no difficulty achieving reduced $\chi^2$ of order unity with no negative emission, and the average time per DEM was approximately $0.13$ milliseconds, or $44$ seconds for the entire active region at full AIA resolution. We find a relatively hot active region core with median temperature of around three million Kelvin, surrounded by cooler emission with median temperature of around two million Kelvin. We also plotted DEMs as a function of length along a coronal loop segment visible in the active region. These DEMs are dominated by a relatively bright background, making it unclear what emission is associated with the loop; background subtraction of the data will be necessary to isolate the loop emission from its surroundings.
\appendix
\section{Equality of SVD-derived and Instrument Response Basis DEMs}\label{app:instresp_demderivation}
The fundamental equation we want to invert is the following:
\begin{equation}
I_j = \int R_j(T)\dem(T)dT
\end{equation}
Without loss of generality, we can represent this integral in matrix form:
\begin{equation}\label{eq:int_eqn_matrixform}
\bm{I} = {\bf R}\bm{E},
\end{equation}
where the elements of the $n \times m$ (for $n_c$ channels and $n_t$ temperature bins) response matrix ${\bf R}$ are $R_j(T_k)$, and the elements of the $n_t$-element DEM vector $\bm{E}$ are $\dem(T_j)\Delta T$. The singular value decomposition of ${\bf R}$ is
\begin{equation}
{\bf R} = {\bf U}{\bf S}{\bf V}^\top,
\end{equation}
Where ${\bf U}$ is an $n_c\times n_c$ unitary matrix, ${\bf V}$ is an $n_t\times n_t$ unitary matrix, and $S$ is the $n_c\times n_t$ diagonal matrix of the singular values of ${\bf R}$. The SVD can be used to estimate $\bm{E}$ by computing the pseudo-inverse of of ${\bf R}$, ${\bf R}^+\equiv {\bf V}{\bf S}^\top{\bf U}^\top$:
\begin{equation}
\bm{E}^{\mathrm{\footnotesize SVD}} = {\bf R}^+\bm{I}.
\end{equation}
We can also choose to represent the DEM using a basis of instrument response functions:
\begin{equation}
\bm{E}^{\mathrm{\footnotesize IR}} = {\bf R}^\top \bm{e}
\end{equation}
The $n_c$-element vector of coefficients $\bm{e}$ is obtained from $\bm{I}$ by plugging $\bm{E}^{\mathrm{\footnotesize IR}}$ into equation \ref{eq:int_eqn_matrixform} and inverting:
\begin{equation}
\bm{e} = [{\bf RR}^\top]^{-1}\bm{I}
\end{equation}
Expanding $[{\bf RR}^\top]^{-1}$ in terms of the SVD of ${\bf R}$, we find that
\begin{equation}
[{\bf RR}^\top]^{-1} = [{\bf U}{\bf S}{\bf V}^\top {\bf V}{\bf S}^\top{\bf U}^\top]^{-1} = [{\bf U}{\bf S}{\bf S}^\top{\bf U}^\top]^{-1} = {\bf U}({\bf S}^+)^\top{\bf S}^+{\bf U}^\top.
\end{equation}
We then obtain
\begin{equation} 
\bm{E}^{\mathrm{\footnotesize IR}} = {\bf R}^\top [{\bf RR}^\top]^{-1}\bm{I} = {\bf V}{\bf S}^\top{\bf U}^\top ({\bf U}({\bf S}^+)^\top{\bf S}^+{\bf U}^\top)\bm{I} = {\bf V}{\bf S}^+{\bf U}^\top\bm{I} = {\bf R}^+\bm{I}.
\end{equation}
Therefore, $\bm{E}^{\mathrm{\footnotesize IR}} = \bm{E}^{\mathrm{\footnotesize SVD}}$, as claimed.
\bibliographystyle{apj}
\bibliography{FastDEMs_jplowman}
\end{document}